\documentclass{article}
\usepackage[utf8]{inputenc}
\usepackage{graphicx}
\usepackage{cite}
\usepackage{multirow}
\usepackage{rotating}
\usepackage[table,xcdraw]{xcolor}

\newcommand\blfootnote[1]{%
  \begingroup
  \renewcommand\thefootnote{}\footnote{#1}%
  \addtocounter{footnote}{-1}%
  \endgroup
}

\title{Engineering and implementation of the Simulation of Automatic Exposure Notification (SimAEN) model}
\author{Gwendolyn Gettliffe, Louise Ivers, Adam Norige, Ted Londner, 
Jonathan Saunders, Dieter Schuldt,Bill Streilein}
\author{
  Gwendolyn Gettliffe, Adam Norige, Ted Londner, \\ Jonathan Saunders, Dieter Schuldt, William Streilein \\
  \\
  MIT Lincoln Laboratory, Lexington, MA\\
  }
\date{13 December 2021}

\begin{document}

\maketitle
\blfootnote{DISTRIBUTION STATEMENT A. Approved for public release. Distribution is unlimited. This material is based upon work supported by the Under Secretary of Defense for Research and Engineering under Air Force ContractNo. FA8702-15-D-0001. Any opinions, findings, conclusions or recommendations expressed in this material are those of the author(s) and do not necessarily reflect the views of the Under Secretary of Defense for Research and Engineering. © 2020 Massachusetts Institute of Technology\\
Delivered to the U.S. Government with Unlimited Rights, as defined in DFARS Part 252.227-7013 or 7014 (Feb 2014). Notwithstanding any copyright notice, U.S. Government rights in this work are defined by DFARS 252.227-7013 or DFARS252.227-7014 as detailed above. Use of this work other than as specifically authorized by the U.S. Government may violate any copyrights that exist in this work.}
\section{Introduction}

Throughout the COVID-19 pandemic, public health institutions have used contact tracing to attempt to slow the spread of the disease. In this scheme contact tracers (individuals hired by the state) identify the \textit{contacts} of known infected people: those individuals who may have been exposed to the disease through their interaction with an infected person. Contact tracers notify these contacts of their potential exposure and encourage them to take mitigation efforts such as self isolation and getting tested with the intention that this will stave off further spread.

The success of contact tracing hinges on several factors, including (1) the ability of public health to identify and communicate with infected individuals, (2) the ability and willingness of infected individuals to share their contacts, (3) the ability of public health to notify these contacts, and (4) the willingness of these contacts to comply with public health instructions. The COVID-19 pandemic has affected these factors in ways that have challenged contact tracing efforts:

\begin{enumerate}
  
  \item In many jurisdictions, the size of the outbreak has exceeded local contact tracing capacity 
  
  \item Infected individuals may have more contacts than they can identify
  
  \item Infected individuals may not know that they are infected
  
  \item Distrust of public health institutions prevents some individuals from cooperating with contact tracing efforts
\end{enumerate}

The specific distance constituting \textit{close enough} and the specific time constituting \textit{long enough} are highly situational and a subject of ongoing study. However, they are often such that an infected individual has more contacts than he or she can identify. In addition, individuals can be infected with and spread SARS-CoV-2 without presenting any symptoms, and therefore might not be aware that they are infected.

The way that COVID-19 spreads has contributed to these challenges. SARS-CoV-2, the virus that causes COVID-19, is spread through airborne droplets exhaled by infectious individuals. An infected individual's contacts include anyone who was close enough to the individual for long en
\textit{Automated exposure notification} (AEN) was devised in March of 2020 \cite{canetti2020} to complement the traditional contact tracing scheme described above, referred to in this paper as \textit{manual contact tracing} (MCT). The overarching goal of AEN is the same as MCT: to notify individuals of their potential exposure to a disease to mitigate the disease's spread. AEN seeks to complement MCT by notifying contacts that MCT may have missed.

AEN uses the Bluetooth technology present on smartphones to estimate when individuals were \textit{too close for too long} (TC4TL) to an infected person and are therefore at risk for infection themselves. These individuals are notified of their exposure through alerts on their smartphones and may include instructions about testing or recommending quarantine. The process is anonymous: notified contacts are not made aware of which infected individual set off the alert, and neither public health nor the infected person is aware of who receives notifications.

What is considered TC4TL is determined by public health institutions, and this judgment is encoded in AEN by setting the sensitivity of the Bluetooth-based assessment. As this sensitivity increases, either the distance corresponding to \textit{too close} increases, the time corresponding to \textit{too long} decreases, or both. The intended effect of increasing sensitivity is to alert more infected individuals, with the goal of slowing disease spread. The side effect is that more uninfected individuals will be unnecessarily alerted. Many of these uninfected individuals will seek a COVID-19 test, potentially burdening their local test infrastructure. They may also quarantine unnecessarily, potentially burdening their local economies. This trade-off between slowing disease spread and increasing public burden leads to a critical question: \textit{How should public health set the sensitivity of AEN?}

The answer depends on a variety of factors, including:

\begin{enumerate}
  \item The properties and prevalence of the disease
  \item The performance of the Bluetooth-based sensor
  \item The behavior of individuals within the jurisdiction
  \item The workflows and capacities of public health
\end{enumerate} 

\noindent These variables interact in complex ways, preventing obvious answers to the question above. 

This paper presents SimAEN,\footnote{SimAEN, pronounced ``SIM-ee-uhn,'' is short for ``Simulation of AEN.''} an agent-based simulation whose purpose is to assist public health in understanding and controlling AEN. SimAEN models a population of interacting individuals (``agents'') in which COVID-19 is spreading. These individuals interact with a public health system that includes AEN and MCT. These interactions influence when individuals enter and leave quarantine, affecting the spread of the simulated disease. Over 70 user-configurable parameters influence the outcome of SimAEN's simulations. These parameters allow the user to tailor SimAEN to a specific public health jurisdiction and to test the effects of various interventions, including different sensitivity settings of AEN.

Public health has to look at not just the benefits of an intervention, but the costs to society as well. If no one ever leaves their houses the disease will be contained, but this presents far too high a burden on society. Prior work has focused on the disease spread half of this problem, identifying the effectiveness of the intervention at reducing transmission. With AEN there are a lot of free parameters for public health to consider so they need a tool that can tell them what those settings should be. If the Bluetooth settings are too tight then AEN may not be sufficiently effective to warrant adoption, while settings that are too loose will result in an unacceptable number of false quarantines. SimAEN provides public health with the insights they need to make these decisions.

\subsection{COVID-19}
COVID-19 is a viral respiratory disease, spread primarily through droplets expelled from the lungs of infectious individuals. Characterization efforts are underway, but based on our current understanding of this and similar diseases there is a large collection of parameters that have to be understood in order to create a complete model of the transmission environment.

One parameter of particular importance for our model is the rate of asymptomatic case generation. Studies have found that between 30\% \cite{mizumoto2020} and 73\% \cite{poletti2020} of cases do not develop symptoms, but are still capable of transmitting the disease \cite{sayampanathan2020}. Being able to curtail this silent transmission stream could be quite beneficial in eliminating the spread of the disease.

Another important aspect of COVID-19 is the method of spread. While there are still questions as to the relative transmission rates between airborne and deposition \cite{smith2020}, it is generally accepted that the virus is only active for on the order of an hour. This limits the number of people who are not in direct contact with an individual who will contract the disease. The spread is also confined, to some extent, making the model different to other diseases (particularly waterborne) that require modeling the channel of transmission \cite{rinaldo2012}.

\subsection{Public Health Response}
The capabilities that public health can employ when respond to the outbreak of a disease are quite limited. Prior to the advent of smart phones these were essentially: recommendations for social distancing and mask wearing, decontamination of surfaces, surveillance testing, and manual contact tracing (MCT).

Social distancing and mask wearing attempt to interrupt the disease spread by preventing the exchange of the contagion between people. The effectiveness of social distancing is limited by the persistence of the contagion in the air, as it doesn't matter how far you stand apart from someone if you subsequently walk through a plume of their previously exhaled breath. Masking attempts to reduce the amount of contagion released into the environment by providing a physical barrier. Further, masks create turbulence which reduces the distance that disease particles travel.

Decontamination of surfaces attempts to remove fomites before they can be inhaled by susceptible individuals. Many diseases remain on surfaces for a protracted period of time and will readily activate upon entering a host. However, in the case of COVID-19 this represents a vanishingly small risk\cite{mondelli2020} and as such is not considered here.

Surveillance testing aims to catch infected individuals early in the disease progression, and get them into quarantine. This curtails the number of people that an infective person interacts with, and thus the number of people they are able to infect. The effectiveness of surveillance testing is controlled by the testing capacity of the system, the ability of individuals to access a testing site, and willingness of individuals to be tested.

\subsection{Manual Contact Tracing}
In the manual contact tracing process, public health reaches out to individuals who have tested positive in order to determine who else they may have come into contact with during the time in which they were infectious. These interviews can take as long as several hours, which makes them an expensive and time consuming process.

The effectiveness of MCT relies on the ability of individuals to remember all of the people that they have been in contact with. Certainly people are able to identify their family and any close friends that they have interacted with, but recall is expected to fall off rapidly as the contacts become less personal, such as people who happened to be at the same store at the same time. Further, the close contacts are likely to be aware of the infection status of the infected individual so there is less value in alerting them through MCT.

The limitations on MCT also include its voluntary nature, which limits who ends up participating. People are unlikely to answer calls from unknown numbers which can make it difficult for public health to alert the potential contacts.
\subsection{Automatic Exposure Notification}

This work focuses on the effects and effectiveness of Automatic Exposure Notification (AEN). AEN is a new capability that is previously untested in a pandemic. It attempts to serve the same role as MCT, but in an automated fashion that is more capable of identifying close contacts. Details of AEN can be found in \cite{chan2020}, but a brief overview is provided in the Background/PACT subsection of this paper.

Prior works \cite{ferretti2020} have found that AEN has the potential to control the spread of COVID-19. Our work augments this literature by creating an agent based model, which is able to be rapidly adapted to additional protocols as they are developed. Our model also has a more realistic testing and quarantine model which takes delays of testing into consideration and allows for transition out of quarantine after negative tests (in addition to other nuances). Further, since this work is focused on the public health response, the model provides probabilities associated with the intricacies of AEN implementation.

\section{Background}
\subsection{Literature Review}
The prediction and estimation of disease spread has been performed using many different methods. In general, these methods land in three broad categories: compartmental models, data-driven models, and agent-based models.
\subsubsection{Compartmental models}
Compartmental models use a series of interdependent ordinary differential equations (ODEs) to describe the mean number of individuals in each of several categories. The total population is broken down into several subpopulations (susceptible, exposed, infectious, and recovered among others) and the fraction of the population in each of the subpopulations changes over time according to the prescribed ODEs \cite{kermack1927}. These models are typically described using a convenient shorthand (SIR, SEIR, etc.) indicating the subpopulations being considered. Owing to their relatively easy evaluation and understandable results, these models have gained wide adoption in the epidemiological community. To gain this tractability the model makes many assumptions, such as that the entire population is homogeneously mixed. However, this assumption and other aspects of their formulation limits their ability to predict the real-world operational changes that occur when resources are limited and interactions are conditionally probabilistic. Small world models, such as those by Strogatz and Watts \cite{watts1998}, show that mixing heterogeneity is omnipresent in human interaction and greatly affects the propagation of disease.
\subsubsection{Data-driven models}
The idea behind data-driven modeling is that given sufficient data about a population you can discover a relationship among the inputs that minimizes the error signal between the model and some set of real world outputs. This can take the form of neural networks \cite{wieczorek2020}, fuzzy logic systems \cite{traulsen2012}, or regressions \cite{werneck2002} which all use prior data to train the weights in the systems of equations that constitute the model. One problem with these models is that they can be opaque, meaning their internal workings are difficult to interpret and their outputs hard to justify.

This sort of model does not consider the population dynamics explicitly, instead focusing on measurable data and assuming that there is an underlying sensible reality generating them. By creating a system with enough freedom, the model is able to work as a surrogate for this reality. However, since they are only a product of the data expert knowledge might not be considered.

Further, since they require initial data and only represent reality as it is currently operating these models are not suited for investigating potential changes in the system or in the early stages of a disease before sufficient data has been collected.
\subsubsection{Agent-based models}
Agent-based models attempt to limit abstraction, instead depicting every member of the susceptible population and adjusting their condition based on the progression of the disease according to a set of rules.
On the most abstract side of agent-based modeling are grid based simulations known as cellular automata. This framework is most famously implemented in Conway’s game of life \cite{adamatzky2010} but has also been applied to disease dynamics \cite{white2007}. These models treat the agents as nodes of a regular grid or irregular network \cite{gagliardi2010} and update the state of these agents based on the state of their neighbors.

Less abstract agent-based models account for spatial variation in interactions. These models create agents that conform to ‘schedules’ as they move through a virtual landscape. The population of agents can mirror reality in terms of demographics, behavior, and spatial distribution. Underlying these models is the assumption that if reality is intricately modeled then the model will behave in the same way as reality.

All of this verisimilitude comes at a cost. The amount of computation required to evaluate the model goes up as a function of the number of agents and memory requirements expand as the list of agent parameters (e.g., features) grows. For these reasons, these models are reasonably recent developments, only just now being fully realizable thanks to modern computing capabilities.

\subsection{PACT}
It was recognized early in the COVID-19 pandemic that contact tracing had the potential to be very important in stemming the spread of the disease. With the advent of smart phones, new potentials for contact tracing have opened up, utilizing the transmission of Bluetooth Low Energy (BLE) chirps to identify proximity. Privacy concerns have always overshadowed contact tracing efforts \cite{levine1988}. While technology can alleviate some of these concerns it also introduces new avenues of attack and exploitation.

\subsubsection{History}
To meet this challenge the Private Automated Contact Tracing (PACT) initiative was formed, joining collaborators from MIT Computer Science and Artificial Intelligence Laboratory (MIT-CSAIL), Boston University, Brown University, Carnegie Mellon University, the MIT Media Lab, the Weizmann Institute, with medical experts at Massachusetts General Hospital Center for Global Health and a number of public and private research and development centers.

The main output of PACT is a privacy preserving protocol that can be employed by smart phone manufacturers to bring the capability of automatic contact tracing to their users. The first public version (v0.1) of the protocol was released in April of 2020 and was quickly implemented by both Apple and Google. Since then governments have been working to integrate this capability into their public health workflows.

PACT has continued to work to address questions about the efficacy and privacy of the protocol. This includes marrying the protocol to the physics of BLE, meeting any newly identified privacy vulnerabilities, working with state public health authorities to determine how best to implement AEN in their state, and ultimately showing how well AEN does in limiting the proliferation of the disease.
\subsubsection{How It Works}
AEN attempts to keep a log of all people what an individual interacts with, including both the length and distance of that interaction. When combined with the weighting scheme this data generates a score for each interaction. Scores for interactions with positive individuals will generate an alert from the app.

When two individuals, each running the AEN smartphone application, are close enough to each other that the Bluetooth signal from one phone is detected by the other a key is exchanged between the phones. This key is stored by the app for future comparison, along with the signal strength associated with the broadcast. These keys are rotated periodically as part of the privacy scheme.

After receiving a positive test an individual uploads to a remote server all of the keys that they have used over some prior period. This results in a database of all keys for all positive individuals. Periodically the app will download this digest of keys and compare them to the keys stored by the app. If there are keys in common between the two sets then the broadcast strength is multiplied by the weight associated with that strength and a sum of these values is generated over all of the matching keys. If the sum is sufficiently high then the individual is alerted of their potential exposure and directs them to subsequent actions that they should take.

A more complete description can be found in \cite{chan2020}

\subsubsection{Health Care}
Identifying nearby individuals is only part of the equation that has to be considered. This capability has to be worked into a public health workflow that controls the access to the information and uses it in conjunction with their other efforts to control disease spread.

Before the system is rolled out the State has to select the weights associated with contact distance and probability. This will determine how many people are sent AEN alerts and subsequently how many people come into contact with PH.

The next PH/AEN touch point is through the testing procedure. Some states have opted to require a call from a PH worker before an app user can upload their keys, while others simply provide the key as part of a positive test result.

Public health also has to select a messaging scheme to send out as part of the AEN alert. Some states may opt to have individuals receiving an AEN alert to call into PH, while others just instruct the recipients to self-isolate and get tested.

The choices that the State makes in regards to these questions will determine the workload that is experienced by their contact tracers and other PH professionals. SimAEN aims to provide guidance to State to help them select the settings and structure their program in the way that best serves their constituents.
\section{BLEMUR}
\subsection{Model Description}
BLEMUR \footnote{Pronounced BLEE-mr}, or Bluetooth Low Energy Model of User Risk, is a model of the probability of alert for contacts of an index case given a specific configuration of AEN and the capabilities and limitations of Bluetooth Low Energy signals as a proxy for distance between people. BLEMUR draws from BLE data taken in MIT Lincoln Laboratory's Autonomous Systems Development Facility (ASDF) at a range of distances between Android phones carried by mannequins that simulate the RF properties of the human body.

BLEMUR's results are also available as probabilities of detection, miss, or false alarm, and a false discovery rate, depending on the user’s definition of an exposed contact. The default definition of an exposed contact is a contact that has been within 6 feet of the index case for 15 minutes or more.
\subsection{Data Collection and Implementation}
BLE signal strength and attenuation data (in dBm) were collected with and without ``body-blocking'' between the phones to capture how having bodies between the phones impacted the attenuation of the signal. Data was also collected with the phone in various locations on the mannequin, including front shirt pocket, front pants pocket, and a messenger bag worn in front of the body.

Histograms were formed of the attenuation data at each of the distances where data was collected (2, 3, 6, 9, 12, 15, 20, 25, and 30 feet), and kernel distributions were fit to the histograms in order to estimate the probability density function (PDF) of attenuation at each collected distance.

For distances that weren't collected, one of the PDFs was used as a template for the PDFs at uncollected distances, shifted so that the median of the new distribution aligned with the interpolated median at that distance.

The PDFs were used within the model to calculate the probabilities that a contact at a given distance would be recorded as each attenuation value. These probabilities combine to determine the probability that a contact at a given distance would be recorded in each of the four BLE attenuation buckets, described in the next section.
\subsection{Weighting Schemes and Buckets}
There are four BLE attenuation buckets in the AEN implementation, the Immediate, Near, Medium, and Other buckets. These buckets are defined by attenuation thresholds, the Immediate, Near, and Medium thresholds. Signals with attenuations below the Immediate threshold would be categorized as within the Immediate bucket, for instance.

Each bucket is assigned a weight between 0 and 2.5. Multiplying these weights by the number of minutes that a contact spends at an attenuation in each bucket, a value of ``weighted minutes-at-attenuation'' (WMs) can be summed up across the buckets. WMs are the metric of exposure risk – the more WMs an individual racks up, the more likely they have been exposed.

The actual implementation of AEN allows WMs to be summed from making contact with multiple index cases to trigger an alert for an individual. Since the BLEMUR model is focused on modeling the contacts of a single index case, our calculation of WMs only considers the contributions of a single index case to each individual contact.

Also, the BLEMUR model does not consider infectiousness of the index case or report type, which are additional parameters that the real AEN system factors in by multiplying WMs by weights for both parameters to get a value of ``exposure minutes'' (EMs). BLEMUR treats WMs as EMs for the purposes of setting a threshold for alert.

Public health authorities decide where to set the threshold of EMs for alert. A low threshold has a higher probability of detected exposed individuals, but comes at the cost of a higher false alarm rate. A high threshold will reduce false alarms but may miss exposed individuals that they would otherwise wish to alert. 
Public health authorities also have the ability to decide where to set the attenuation thresholds and weights for the buckets. As a starting point, two sets of parameters were proposed by the Risk Score Symposium Invitational (``RSSI'') Workshop : a ``Narrower Net'', which prioritizes a lower false alarm rate but at a higher miss rate, and a “Wider Net”, which prioritizes a lower miss rate at the expense of a higher false alarm rate. The settings are included in the table below.

% \usepackage{multirow}
\begin{table}[]
\begin{tabular}{|l|l|l|l|l|l|}
\hline
         & & \textbf{Immediate} & \textbf{Near} & \textbf{Medium} & \textbf{Other} \\ \hline
\multirow{2}{*}{} \textbf{Narrower Net} & Threshold & 55dB & 63dB & 70dB & - \\ \cline{2-6}
         & Weight & 150\% & 100\% & 40\% & 0\% \\ \hline
\multirow{2}{*}{} \textbf{Wider} Net & Threshold & 55dB & 70db & 80db & - \\ \cline{2-6}
         & Weight & 200\% & 100\% & 25\% & 0\% \\ \hline
\end{tabular}
\end{table}

The calculation of contact duration is based on the app ``listening'' for Bluetooth signals once every five minutes. BLEMUR assumes that each sample equals five minutes of contact with the index case. This sampling rate, depending on when the contact begins relative to the last sample, can lead to contact durations being underestimated.

BLEMUR determines the probability that a contact of an $x$ minute duration is recorded as either $x$ or $x-5$ minutes, and this probability factors into the probability that a contact of an actual proximity and duration accumulates enough WMs to cross the alert threshold.
\subsection{Inputs and Outputs}
\underline{Input: Contact distribution}

BLEMUR accepts a list of $N$ contacts, given in pairs of distance and duration. For the purposes of this model, the contacts are assumed to stay a fixed distance from the index case for the duration of the contact period. Two contact models exist:
\begin{enumerate}
\item	Contact grid: The contact grid is a list of 900 contacts, one at each point in a grid of 1-30' in distance and 1-30 minutes in duration. This is not a very “realistic” contact distribution for the average index case, but it provides good visualization of which regions within the distance-domain are most likely to be alerted, either falsely or correctly. This can help public health authorities visually understand what types of contacts are likely to be missed or falsely alerted.

\item	Representative contact model: The representative model creates a list of contacts drawn from distributions defined by the index case behavior (are they working from home or going to public places like a grocery store) and U.S. household size statistics. Multiple draws can be made to form conclusions on the range of potential probabilities of detection and false alarms.
\end{enumerate}

\noindent\underline{Input: Weighting schemes and buckets}

The weights and thresholds separating attenuation buckets are described in an earlier section. BLEMUR allows users to easily change these weights and thresholds to see how the outputs are affected.

\noindent\underline{Output: Classification}

The primary output of BLEMUR is a list of:
\begin{itemize}
\item probability of detection
\item probability of false alarm
\item false detection rate
\end{itemize}

These rates are highly dependent on the contact distribution assumed, so it is important that any discussion of detection and false alarm rates be accompanied by a description of the contact distribution used. Pd and FDR are inputs to SimAEN. An optional output are the expected counts of contacts for the input contact distribution for each of the rates mentioned above. 

\section{SimAEN}
\subsection{Model Description}
The SimAEN model definition is based on the ODD protocol as specified in \cite{grimm2010}. This formulation establishes a standard way of communicating an agent-based model in a manner that allows for re-creation by other interested parties. The agents in this description represent people and we will refer to them as agents, people, or individuals interchangeably.
\subsubsection{Overview}
\paragraph{Purpose}
AEN is a new technology and there is need to understand its effectiveness. The SimAEN agent-based simulation was developed to explore and understand how AEN and its internal configuration and components affect disease spread. In addition to the independent effects of AEN we also endeavor to show the effects of manual contact tracing, widespread testing, and the results of deploying these strategies in combination with each other.
\paragraph{Entities, state variables, and scales}
Agents in the SimAEN model represent individuals whose interactions are guided by a collection of probabilities. These individuals operate within the context of a ``world'' which also has a collection of probabilities and parameters assigned to various aspects of it (e.g, the duration of each phase of the disease, the probability of a call from public health being answered by an individual, etc.). Agent-based models are naturally suited to object-oriented programming methods, so both individuals and worlds can be thought of as objects -- though for this simulation only a single world exists at any one time.

Further information on the parameters of the individual and world objects can be found in the appendix. These parameters describe the various states that individuals can be in during the simulation.

The world in which individuals exist advances on a discrete schedule, where state changes occur once per day. This time frame (e.g., once per day) was chosen because it captures the phenomena relevant to disease transmission, while also being long enough that it does not take an unacceptable amount of computation to determine results. In addition, longer time frames, such as (e.g., once per week) would not permit the granularity associated with testing events or other characteristics of interest.

The individual is the smallest agent level considered. This model does not directly consider family units or workplace structures. The type of transmission events that take place in these settings are accounted for by modeling the events using a log-normal distribution. This distribution features a long tail, meaning that there is a relatively high probability of an event occurring where a large number of individuals contract the disease (e.g., in the tail of the distribution).
\subsubsection{Process overview and scheduling}
The simulation advances one day at a time, during which each of the individuals in the simulation updates its status, potentially gets tested, is processed by public health, and changes its behavior.

Each day is broken down into a series of events where various aspects of the simulation are performed. These events always occur in the specified order, though this order was chosen arbitrarily. 

The first event to be processed is transmission. During this event each infected individual is evaluated to see if they produced additional infections. Some number of uninfected individuals are also produced, based on the false detection rate (FDR) and the probability of transmission associated with the individual.

Following the transmission event comes the testing event during which all individuals are checked to see if they get a test based on a probability conditioned on the agent's traits. If a test is performed then a countdown is started, simulating the delay that occurs between test and results. The test is also entered into a queue to account for potential limits on the testing capacity of public health.

The next event is automatic tracing, where notifications are sent out by individuals who have tested positive and have probabilistically decided to upload their keys to the AEN server. Receiving a notification from the AEN system will change the individual's probability of getting tested and also altering their behavior.

In the final event, public health performs manual contact tracing. This step involves contacting a person who has tested positive to identify individuals they may have come in contact with and have potentially infected.

At the beginning and end of each day various bookkeeping tasks are executed. This includes removing individuals from the simulation if they have existed for four disease generations, since there is no way for these individuals to pertain to the current state and propagation of the simulation. This saves on memory and future computational cost.
\subsubsection{Design concepts}
\paragraph{Basic principles}
This SimAEN model is based on the transmission and public health response associated with COVID-19. This means that transmission occurs through close contact between individuals, not through contact with a previously exposed surface, consumption of contaminated food or drink, or other methods of disease transfer. This affects the number of people that can be expected to become infected by an individual in any particular transmission event and the likelihood that a person would be able to identify the person who infected them or whom they may have infected.

One important assumption of our model is that the disease only ever infects a small portion of the population. In the standard SEIR model \cite{anderson1979} the rate of change of the susceptible population (i.e., $dS$) is a function of the infected population as a proportion of the overall population. However, as long as this fraction is small we can treat the susceptible population as a constant; a pool of individuals that we can always draw from. As such, we do not model the greater population of people who have not been directly affected by the disease. Instead, individuals are created at the time that they are needed and then disposed of once they are no longer integral to the simulation. The downside of this method is that it allows the number of people affected to grow without bound over long time scales. However, it saves significant computation by not simulating all of the people who are not impacted (which, per our assumption, is most of them). This also allows us to ignore the spatial aspects of individuals since transmission events do not have to occur at a given intersection of modeled individuals’ journeys.

The nature of COVID-19 also determines ranges of potential parameters for stages and duration of the disease. Infections are assumed to last 17 days \cite{peirlinck2020} from initial exposure to recovery. In SimAEN individuals who are infected progress through the following stages of the disease: \texttt{EXPOSED}, \texttt{PRESYMPTOMATIC}, \texttt{SYMPTOMATIC} (or \texttt{ASYMPTOMATIC}), and \texttt{RECOVERED}. The SimAEN model does not account for the difference between recovery and death as both of these outcomes remove the individual from the system and do not have an impact on the methods of mitigation employed by public health.

Testing capability is also modeled based on what has been seen in the COVID-19 pandemic. Evidence from the current testing regimen suggests that there are very few false positive results. There is, however, a relatively high rate of false negatives. These rates are dependent on what stage of disease the individual is in at the time that the test is performed.
\paragraph{Emergence}
We are interested in the proportion of individuals who need to be running the AEN application for their to be a notable effect on disease propagation. This fraction will be dependent on the expected time between when the individual contracting the disease and any of the people they infected receiving an alert on the app. It will also be affected by the other mitigation efforts being taken as they will have potential overlap with the population using AEN.

A quick analysis shows that there is a quadratic effect between the fraction of people who are running the app and the fraction of people who will receive an alert. Assume that the probability of an individual running the app is $x$. Since it requires that both members of an interaction be running the app for either of them to receive an alert this means that an interaction has a probability of $x*x=x^2$ of meeting this criteria.

The other emergent aspect of this model is the feedback between MCT and AEN. Individuals being contacted by MCT are more likely to get tested, and if they are running the app and test positive then more potentially infected individuals will be notified of this fact. This works in the other direction as well. However, there is also the overlap between the two (individuals receiving both an AEN alert and being contacted through MCT) reducing their overall effectiveness.

\paragraph{Adaptation}
The SimAEN model assumes that there are two distinct levels of interaction: \texttt{NORMAL} and \texttt{RESTRICTED}. Agent in the \texttt{RESTRICTED} state have fewer close contacts, leading to lower levels of disease spread. Each day agents are checked to see if they transition to \texttt{RESTRICTED} as a result of their condition. The probability of this transition is conditioned on several traits (such as receiving a positive test or being contacted by public health) that may also change during the course of the simulation.

Agents will also probabilistically don masks, conditioned on their level of interaction. It is assumed that as agents isolate themselves they are also more likely to take other precautions. Once an agent starts wearing a mask it will not stop wearing a mask.

The SimAEN model does not adapt agent behavior based on the progression of time or the prevalence of the disease. That is, agents do not respond to high disease rates by altering their levels of interaction or deciding to wear masks. Changes resulting from public health messaging such as deciding to promote mask wearing are also not modeled.

\subsubsection{Objectives}
Since the adaptation in SimAEN is purely driven by probabilities, there is no objective that is trying to be optimized. In the abstract sense the agents are trying to minimize the amount of exposure to others, but this is projected on them by the selection of probabilities for behavior transition.
\paragraph{Sensing}
Agents are aware only of themselves with the exception that they are able to identify some fraction of the people who they have interacted with for the purposes of contact tracing by public health. Agents are able to identify if they are \texttt{SYMPTOMATIC}, as this is a trait that will affect their probability of getting tested or changing their behavior.
\paragraph{Interaction}
Agents are created along with the set of all individuals who they will ever interact with (apart from the agent that infected them). Each day a subset of this set is randomly selected for interaction, making them eligible for infection.

During automated and manual contact tracing individuals interact through an intermediary, either the AEN application or the public health system. 
\paragraph{Stochasticity}
This model is driven significantly by stochasticity associated with the initial parameter settings. These probabilities are outlined in the appendix, but a brief overview is given here.

How an agent experiences the disease are determined by the latent and incubation times, which are drawn randomly from a normal distribution on a per-agent basis.

Several agent traits have to be set during agent creation. First, the agent is probabilistically set to be wearing a mask. The mask wearing state of the new agent is combined with the mask state of the infecting agent to determine whether the new agent is set as infected or uninfected. Finally, the new agent has some probability that they had previously downloaded and are running the AEN application. The agent is also assigned a collection of individuals who they have the potential to interact with over the course of their time in the simulation.

During transmission events the number of individuals that an agent interacts with is probabilistic, conditioned on the behavior state of the transmitting individual. If the transmitting individual is running the app there is also the probability that they will interact with some number of uninfected agents who were not close enough to have been infected but were identified by the AEN application. This false detection probability models inaccuracies Bluetooth detection capability of digital contact tracing. If the person transmitting the disease and the person being infected are both running the app there is some probability that the app on either end of the transmission will detect the signal.

Whether an agent gets a test is probabilistic, conditioned on whether they have been contacted by public health, received an notification from AEN, tested positive, or are feeling symptomatic.

The probability of a test coming back positive is based on the stage of the disease the agent is in at the time of the test. As noted in the assumptions, there are no false positive test results.

There is a probability that after receiving a positive test an individual will contact public health for the purposes of contact tracing. During contact tracing there is a probability that any given call will successfully reach the agent. If the return call from public health is successful then there is a probability that the traced person infected will be identified.

\begin{figure}[htbp]
 \includegraphics[width=\linewidth]{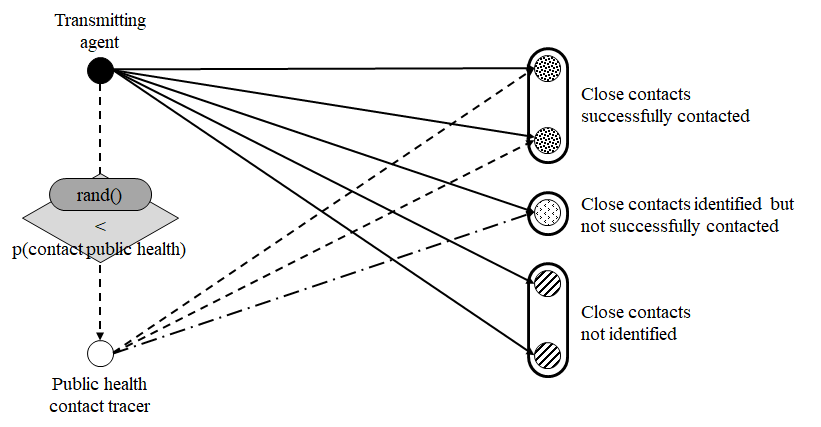}
 \caption{Diagram of agent generation showing the conditions that agents can obtain based on whether they are identified by the generator}
 \label{fig:detection}
\end{figure}

A positive test also comes with the probability that the individual will upload their key to the AEN server. Uploading a key initiates the AEN, alerting all of their contacts who were detected by app.

Finally, there are probabilities associated with an individual changing their interaction behavior, conditioned on the traits of the individual. Agents who transition to \texttt{RESTRICTED} will stay in that condition, except if they receive a negative test, in which case they will probabilistically transition to a behavior state based on the distribution of states currently occupied across the simulation.
\paragraph{Collectives}
There are no collectives, such as families or work groups, included in this model. All individuals are treated based only on their own traits and do not have and probabilities or parameters based on their generator or the agents they are generated alongside.

The closest aspect to a collective in this model is the set of individuals which an agent is generated with. This group represents the entire population of people who that agent has any chance of interacting with. Individuals in this set are the only people that the agent can infect.
\paragraph{Observation}
Each agent keeps track of every transition that it makes while being simulated. Tracking this information supports forensic analysis of the simulation and how it progress. Examples of information that agents track include all of their traits (such as whether they have been tested) along with the days (since simulation start) on which those traits changed. They also keep track of all generation events including the status of all of the people involved in that event. Agents also track things for which they do not have direct knowledge. For example, agents keep track of how many times they have been unsuccessfully called by public health. This information is not used by the agent but supports deeper understanding of the simulation AEN process.
\subsubsection{Details}
\paragraph{Initialization}
The simulation begins with a collection of infected individuals. For simulations starting from the ``initial outbreak'' state we chose 20 infected agents to start. This is a small enough number that it will not overwhelm the steady state yet is large enough that the disease will be able to take hold and propagate. It is assumed that there has been some low level of the disease circulating in the population prior to the exponential grown segment of the spread, and so these 20 individuals are initiated at a random point in the disease progression. This is accomplished by starting them with a “day in system” variable set to a random value drawn from a uniform distribution over the 14 days of the disease lifespan. Each run starts with a new random specification of these 20 agents. These individuals are also assumed to not have been tested, since we are starting from a time before widespread testing is available and AEN is in use.

For simulations of the disease in later states of spread a multiplier was used to account for the potentially large number of individuals being infected. One simulation we performed looked at the state of Massachusetts in February 2021. At this point in time there were an estimated 72270 active cases in the Commonwealth \cite{dong2020}. However, since the exact mix of individuals (what state of the disease they are in) was unknown we had to adjust this number slightly in order to get day-to-day new case numbers which match the observed rates. We found that a starting cases count of 53000 produced this match. This lower number is likely because of lower infectiousness of individuals late in the disease progression. Instead of simulating all 53000 individuals we started with 530 and multiplied all outputs by 100. Simulating only a fraction of the population also requires a change to the number of contact tracers and daily testing capacity since each agent now effectively represents 100. This method slightly decreases the fidelity of the simulation, but the gain in speed is fair compensation.

\paragraph{Input data}
All information necessary for progression of the model is contained within the simulation itself. The only outside input is through the selection of the parameters that drive the system. Once the model is started it carries forward without any additional input.

\subsubsection{Submodels}
The SimAEN model includes submodels for testing, behavior, and public health interventions (ACT and MCT).

The testing submodel assumes that individuals will seek out a test contingent on their traits. Each day a random draw occurs for each individual to determine if they will get a test performed. When an individual gets a test it generates a \texttt{TEST} object. Each day a number of tests is processed, based on the testing capacity parameter. When a test is returned a random draw is made to determine the result (positive or negative), conditioned on the state the agent was in when they were tested. A positive test will further prompt a random determination of whether they will upload their key to the AEN system and/or call public health. Contacting public health will add them to a list of index cases.
 
\begin{figure}[htbp]
 \includegraphics[width=\linewidth]{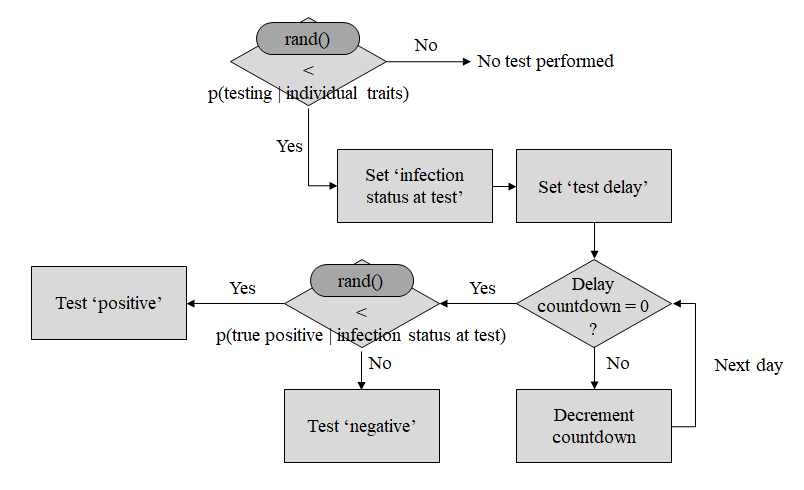}
 \caption{Diagram of testing methodology as implemented in SimAEN}
 \label{fig:testing}
\end{figure}

Behavioral changes of the individual are based on their traits. All agents start out in the normal – or base state – where they are interacting with society in the way that they would have were they not infected. Each day a random draw is made to determine if the individual will transition to the \texttt{RESTRICTED} behavior. It is assumed that an individual in the \texttt{RESTRICTED} state will stay stay in that state unless they are not symptomatic and receive a negative test result. While people in the real world may have day-to-day-variation in their level of interaction we assume that the probabilistic nature of transmission employed in this model is sufficient to describe this and that variation will not significantly affect the measures of interest being studied.

In the AEN submodel individuals receiving a positive test have a random probability of uploading their keys to the AEN system. Some jurisdictions have selected a workflow that requires individuals talk to public health before uploading their key. In this case the individual running the app who received a positive test will also generate a \texttt{CALL} object. Once this call is processed the agent will upload their key. Uploading a key triggers the AEN system to transmit an alert to all close contacts of the uploader who were running the AEN application and received a beacon message during the transmission event. All of these agents will update their object variables to note that a notification was received.

Manual contact tracing (MCT) encompasses both contact tracing and notification of identified contacts. It is assumed that there is a limited number of contact tracers and that each works some period of time each day. Each call they make takes some amount of time, which is based on whether it is a call for the purposes of contact tracing or just as a notification. There is also some time associated with missed calls. Our manual contact tracing submodel processes an ordered list of \texttt{CALL} objects (the “call list”). During the manual contact tracing portion of the \texttt{Main()} (e.g. daily) simulation evaluation loop calls go out to people in the order they were placed on the call list. When a call occurs the appropriate amount of time is subtracted from the available call time:

\vspace{5pt}
$available\:call\:time\:=\:\#\:contact\:tracers\:\cdot\#\:work\:hours/day$
\vspace{5pt}

If the call being made is to an index case then a contact trace is performed. If any individuals are identified during the trace they are added to the call list. Individuals who are not successfully called are added to the back of the contact list. Missed calls are logged and after an individual misses their allotment of calls public health will assume that they are unreachable and remove them from the list. When there are large numbers of people on the call list the call list may not be cleared on a single day. In that case the contact tracing calls will pick up the next day at the point where they left off the day before.

\subsection{Validation}
\begin{figure}[htbp]
 \includegraphics[width=\linewidth]{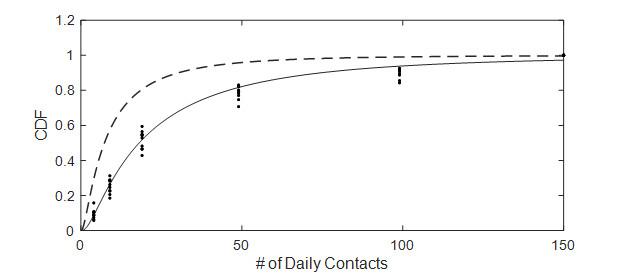}
 \caption{Distribution of daily interactions\cite{fu2005} (dots), along with best fit log normal distribution (solid line; $\mu = 2.1, \sigma = 1.1$) and Pennsylvania fit (dashed line; $\mu = 1.9, \sigma = 1.1$)}
 \label{fig:contacts}
\end{figure}

The parameters used in this paper are based on established research, where available. This includes the probability of transmission for presymptomatic, symptomatic, and asymptomatic individuals\cite{sayampanathan2020}, the rate of asymptomatic cases\cite{mizumoto2020}, the effectiveness of masks\cite{aydin2020}\cite{ueki2020}, the lengths of the incubation \cite{backer2020}, latent\cite{peirlinck2020} and infectious periods\cite{peirlinck2020}, probability of a receiving a positive test \cite{watson2020}, 

Transmission also depends on the number of interactions that an individual has on any given day. Our model assumes that each day an individual interacts with a number of people drawn from a log normal distribution parameterized by $\sigma, \mu$ as shown in \ref{fig:contacts}

Validation of the model was performed by showing that given these parameters the model reproduces the rate of spread in real world conditions. For our case we compared the output of the model with the early days of the outbreak in Pennsylvania. Using the initial phase of the outbreak eliminates several confounding factors such as the prevalence of mask usage, and changes in interaction behavior.

\begin{figure}[htbp]
 \includegraphics[width=\linewidth]{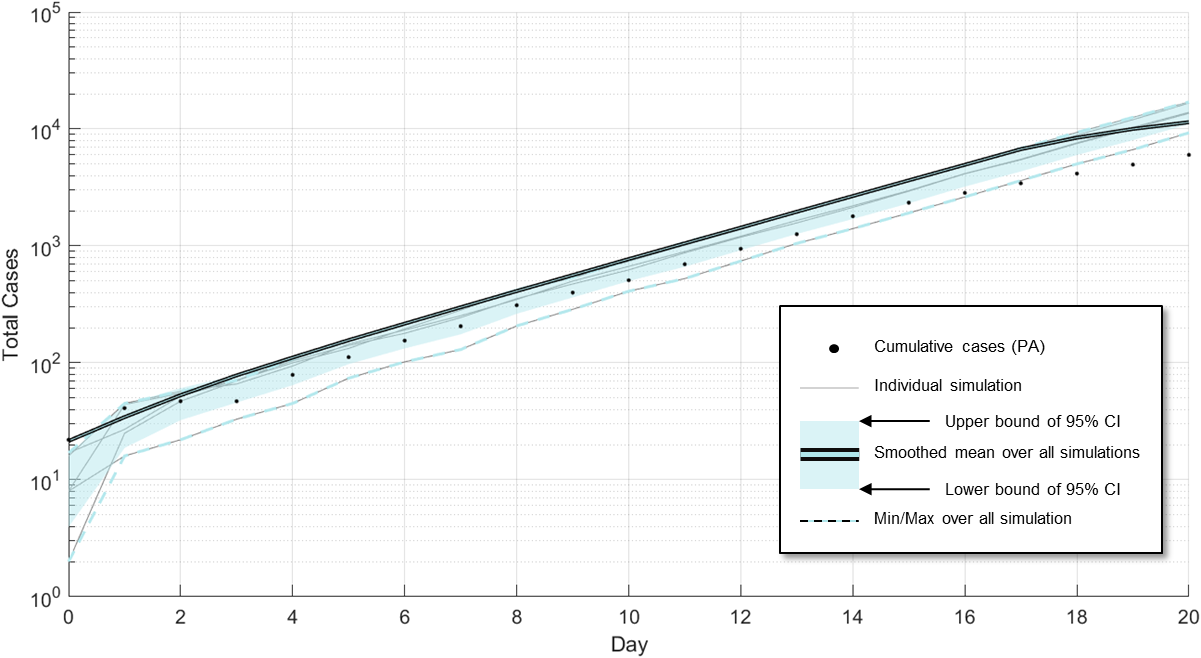}
 \caption{Infection growth for the first 20 days of the COVID-19 outbreak in Pennsylvania along with the results of simulation from SimAEN without any active interventions}
 \label{fig:pennsylvania}
\end{figure}

As can be seen in \ref{fig:pennsylvania} the parameter set shown in \ref{fig:contacts} produces infection curves which match the real life data as observed in Pennsylvania. The use of a mean interaction parameter lower than the best fit curve for \cite{fu2005} ($\mu = 2.1$ vs. $\mu = 1.9$) is justified on the following bases:
\begin{itemize}
	\item The counts from \cite{fu2005} are not unique interactions. Since most people interact with their family and coworkers regularly we would anticipate a lower mean number of unique contacts
	\item The Pennsylvania case counts are for the entire state which includes many rural areas which may have significantly lower numbers of interactions than those found in the more urban study
	\item Due to the high number of asymptomatic cases, the number of reported cases is likely much lower than the true number of cases
	\item The first cases in Pennsylvania (March 10, 2020) were several months after reports from Wuhan (January 2020), which may have prompted changes in behavior of the population
\end{itemize}

\begin{figure}[htbp]
 \includegraphics[width=\linewidth]{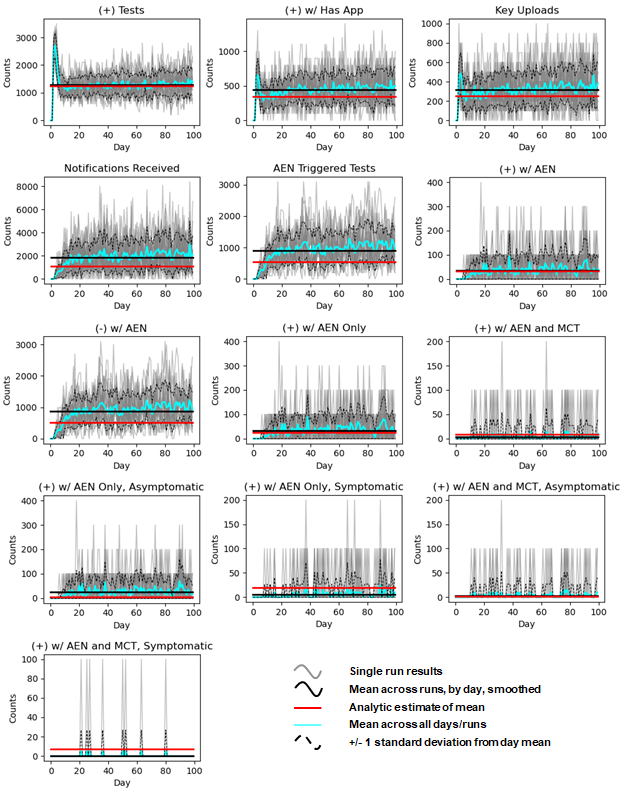}
 \caption{Comparison of SimAEN results and the analytic Massachusetts results}
 \label{fig:rafi}
\end{figure}

The second stage of validation was against an analytic exploration of Massachusetts numbers based on fractions derived from a variety of sources. These are seen in Figure \ref{fig:rafi}, showing that the behavior of the simulation across a wide range of measures is close to the analytic values.

To create \ref{fig:rafi} the simulation parameters were set to observed and published values and run 20 times, each with a different random seed. The results of these runs are shown in grey. The daily mean across these runs is shown as the solid black line. One standard deviation is denoted with dashed black lines. Averaging the daily means produces the overall mean which is shown in cyan. Finally, the analytic result is the solid red line. We see good agreement between the cyan (average simulation) and red (analytic) lines, indicating that the implementation is consistent with what is observed from end to end.

For AEN our parameters (probability of one phone detecting another, and the rate of individuals outside of too close for too long receiving an alert) are based on the experimentation done and processed through BLEMUR.

All other parameters are personal behavior choices which serve as free variables for exploration. This is where SimAEN permits public health to understand the effects of their choices and guide their decision making process.
\subsection{Scope}
The intention of this model is to provide public health officials with the ability to understand the effects of the actions that they have at their disposal. In order to accommodate a large range of potential input parameter values it was critical that the model be optimized for fast execution. To accomplish this we sacrificed some fidelity of the actual spread of the disease, instead focusing on ensuring that the infection rate is properly affected by the public health controllable parameters. The speed gained by not incorporating factors such as mobility data or interaction topology allows the model to be executed numerous times to build up a statistical understanding of the range of potential outputs and to explore more of the space.

This model is also intended for long term understanding, not near term high accuracy estimates of cases or deaths. Assumptions about the predictability or at least stability of population mobility may be reasonable for a 2 week horizon, but decisions made by public health are typically intended to be carried out on the order of months. Decisions about the hiring of contact tracing personnel or the setting of AEN weights are not going to change in response to small fluctuations in infection rate.

\section{Results}
\subsection{Effects of App Adoption}

\begin{figure}[htbp]
 \includegraphics[width=\linewidth]{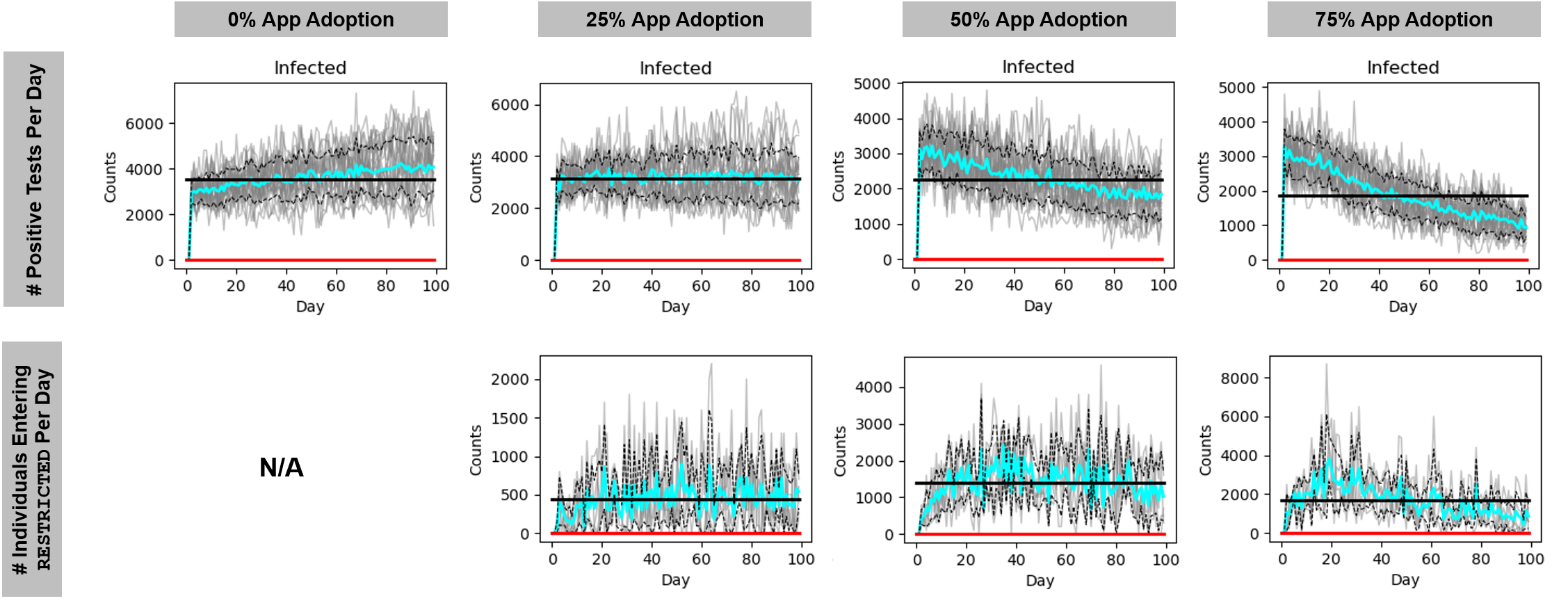}
 \caption{As app adoption increases there is some decrease in the infection rate, but it comes at the cost of unnecessary restriction of individual behavior. Here the interaction rate is 2.1}
 \label{fig:costofapp21}
\end{figure}

The calculated effective reproduction rate (the mean number of new infected per infected individual) for this set of app adoption rates are:

\begin{table}[]
\begin{tabular}{ll|l|l|l|l|}
\cline{3-6}
$R_{eff}$                                                                                      &                 & \multicolumn{4}{l|}{\cellcolor[HTML]{C0C0C0}\textbf{App Adoption Rate}}                                                  \\ \cline{3-6} 
& \textbf{}                   & \cellcolor[HTML]{C0C0C0}0\% & \cellcolor[HTML]{C0C0C0}25\% & \cellcolor[HTML]{C0C0C0}50\% & \cellcolor[HTML]{C0C0C0}75\% \\ \hline
\multicolumn{1}{|l|}{\cellcolor[HTML]{C0C0C0}}                                            & \cellcolor[HTML]{C0C0C0}2.1 & 1.02                        & 0.98                         & 0.94                         & 0.89                         \\ \cline{2-6} 
\multicolumn{1}{|l|}{\cellcolor[HTML]{C0C0C0}}                                            & \cellcolor[HTML]{C0C0C0}2.5 & 1.16                        & 1.13                         & 1.07                         & 1.03                         \\ \cline{2-6} 
\multicolumn{1}{|l|}{\cellcolor[HTML]{C0C0C0}}                                            & \cellcolor[HTML]{C0C0C0}2.7 & 1.21                        & 1.18                         & 1.12                         & 1.07                         \\ \cline{2-6} 
\multicolumn{1}{|l|}{\multirow{-4}{*}{\cellcolor[HTML]{C0C0C0}\textbf{Interaction Rate}}} & \cellcolor[HTML]{C0C0C0}2.9 & 1.27                        & 1.22                         & 1.15                         & 1.08                         \\ \hline
\end{tabular}
\caption{\label{tab:ReffChange}Effects of app adoption across a range of interaction rates. Here we see larger absolute effects on $R_{eff}$ for higher interaction rates}
\end{table}

\begin{figure}[htbp]
 \includegraphics[width=\linewidth]{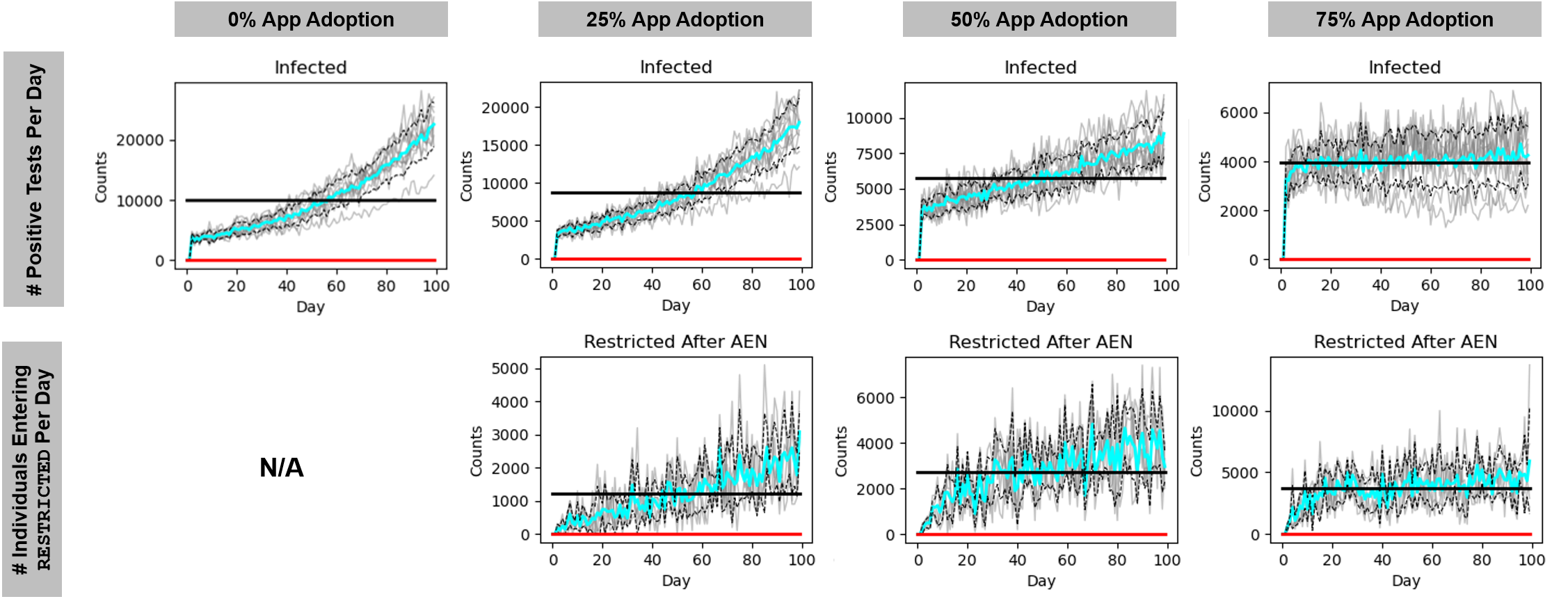}
 \caption{Costs and effects for a slightly elevated interaction rate of 2.5}
 \label{fig:costofapp25}
\end{figure}

\begin{figure}[htbp]
 \includegraphics[width=\linewidth]{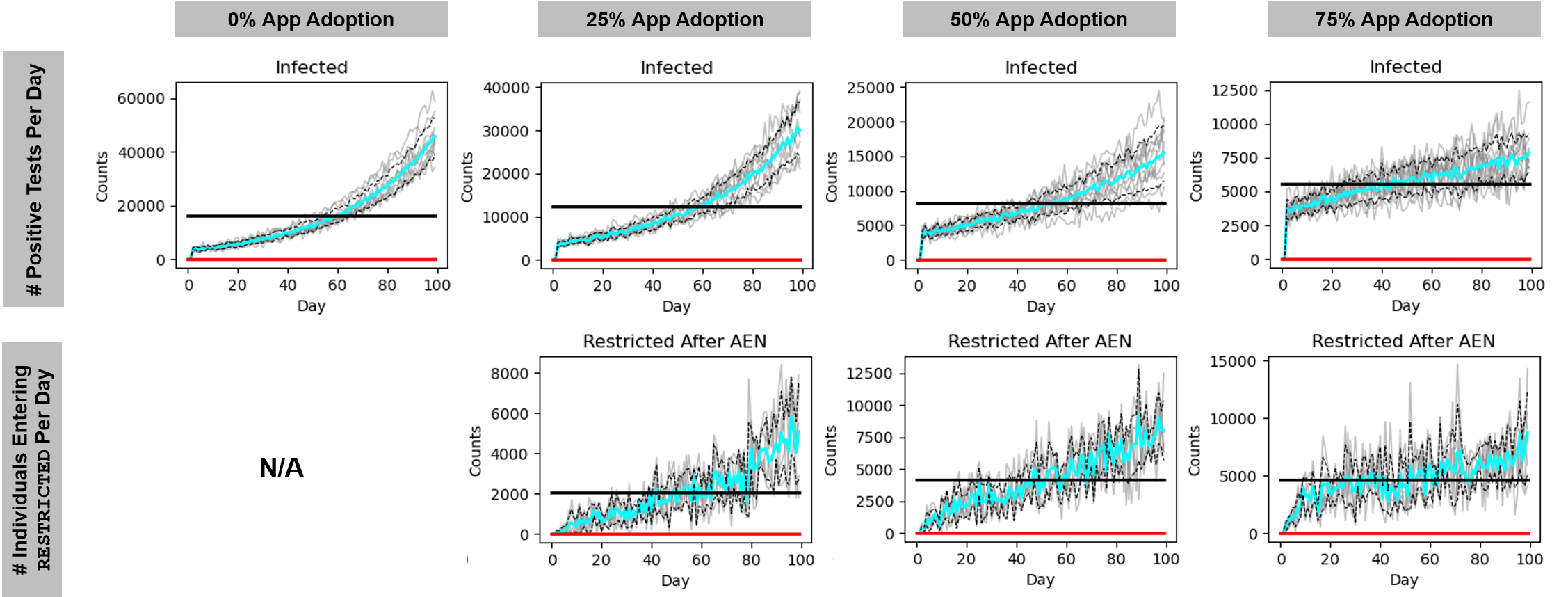}
 \caption{Costs and effects for an interaction rate of 2.7}
 \label{fig:costofapp27}
\end{figure}

\begin{figure}[htbp]
 \includegraphics[width=\linewidth]{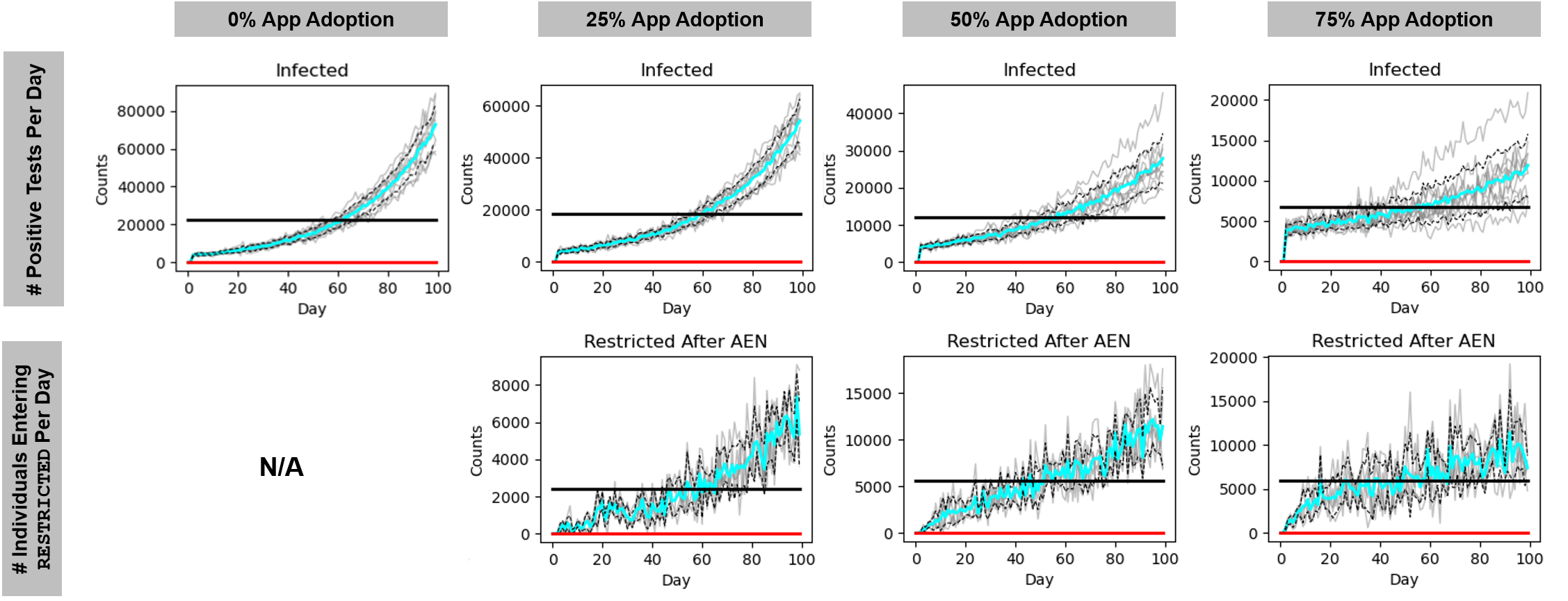}
 \caption{At the interaction rate observed prior to the pandemic ($\mu=2.9$) we see a larger absolute change in $R_{eff}$ than at the lower pandemic level interaction rate.}
 \label{fig:costofapp29}
\end{figure}

In fact, if we look at the percentage reduction in $R_(eff)$ for each level of interaction we see that increases in app adoption are generally equally efficacious. That is, a 50\% app adoption rate will reduce effective reproduction rate by ~7\% regardless of the interaction rate between individuals. This is a result of the increased interaction rate affecting both the number of transmissions but also the number of detections via the app.

\begin{table}[]
\begin{tabular}{ll|l|l|l|l|}
\cline{3-6}
\% Change in $R_{eff}$                                                                         &                 & \multicolumn{4}{l|}{\cellcolor[HTML]{C0C0C0}\textbf{App Adoption Rate}}                                                  \\ \cline{3-6} 
& \textbf{}                   & \cellcolor[HTML]{C0C0C0}0\% & \cellcolor[HTML]{C0C0C0}25\% & \cellcolor[HTML]{C0C0C0}50\% & \cellcolor[HTML]{C0C0C0}75\% \\ \hline
\multicolumn{1}{|l|}{\cellcolor[HTML]{C0C0C0}}                                            & \cellcolor[HTML]{C0C0C0}2.1 & ---                         & 3.8                          & 7.6                          & 12.8                         \\ \cline{2-6} 
\multicolumn{1}{|l|}{\cellcolor[HTML]{C0C0C0}}                                            & \cellcolor[HTML]{C0C0C0}2.5 & ---                         & 2.5                          & 7.6                          & 11.3                         \\ \cline{2-6} 
\multicolumn{1}{|l|}{\cellcolor[HTML]{C0C0C0}}                                            & \cellcolor[HTML]{C0C0C0}2.7 & ---                         & 2.8                          & 8.1                          & 11.7                         \\ \cline{2-6} 
\multicolumn{1}{|l|}{\multirow{-4}{*}{\cellcolor[HTML]{C0C0C0}\textbf{Interaction Rate}}} & \cellcolor[HTML]{C0C0C0}2.9 & ---                         & 4.2                          & 9.5                          & 15.5                         \\ \hline
\end{tabular}
\caption{\label{tab:percentChange}For lower interaction levels the percent change is essentially equivalent. At the highest interaction level the percent change is slightly higher. This is due in part to the app based detections outpacing the effects of MCT}
\end{table}

\subsection{Variations in Detection}
One of the factors that public health has control over is the setting on the app. These weightings determine how likely it is that a person will receive an alert. The more restrictive the settings, the fewer people are identified, but as the focus is loosened more people are quarantined unnecessarily.
\begin{figure}[htbp]
 \includegraphics[width=\linewidth]{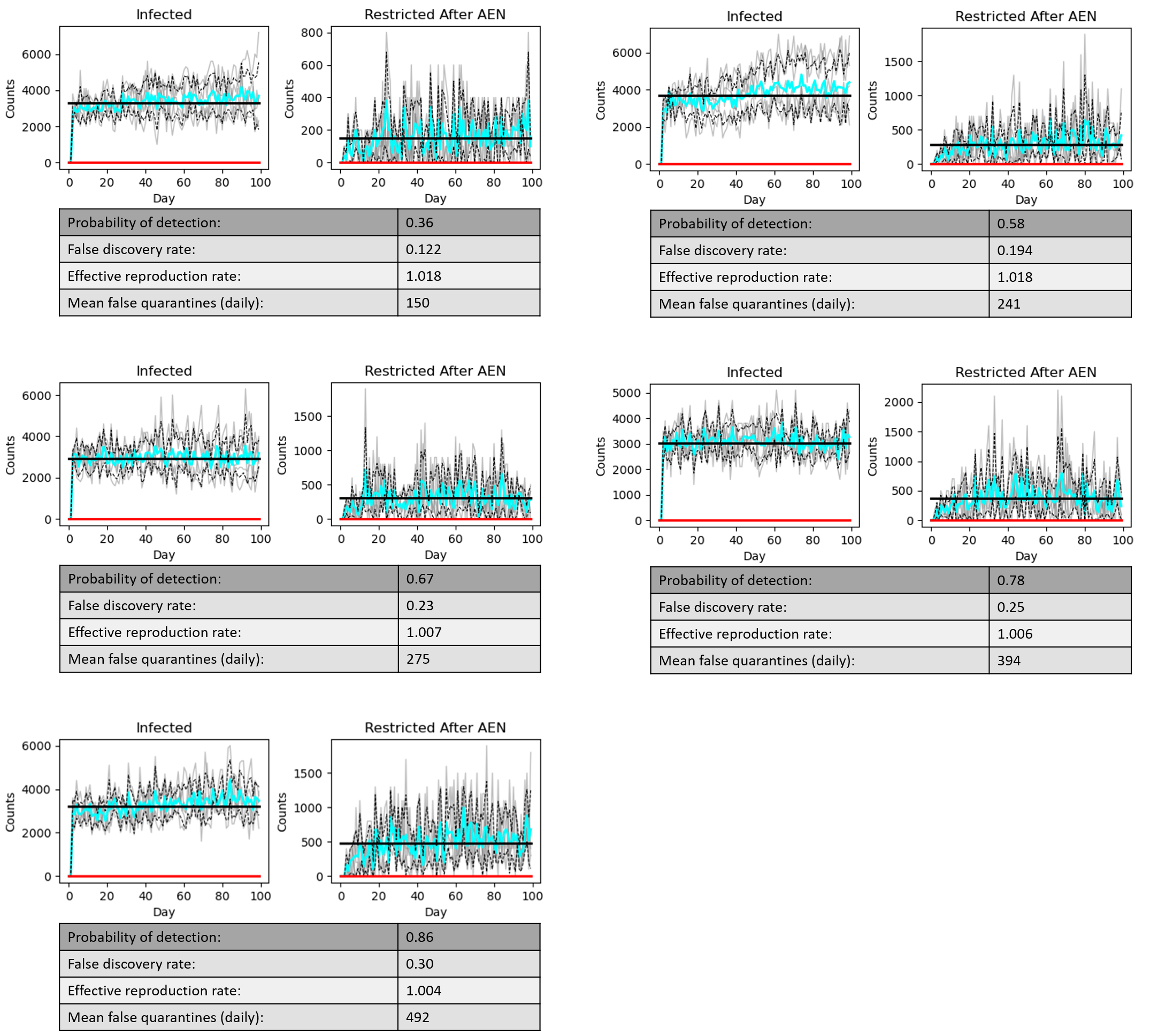}
 \caption{The number of new daily infections along with the number of false quarantines for a collection of probability of detection and false discovery rate}
 \label{fig:pdpfa}
\end{figure}

As can be seen in \ref{fig:pdpfa} there is very little effect on the reproduction rate (for the baseline configuration) as the weights are made less restrictive. On the least restrictive side the reproduction rate is 1.018, while at the most restrictive it is only reduced to 1.004. However, the number of daily false quarantined increased from 150 per day to 492 per day. 
\section{Conclusion}
We have presented an agent-based model of the effects of AEN and public health on the progression of the COVID-19 disease. As demonstrated in the experiments section, the proposed model produces a good fit for the early pandemic data, as measured in Pennsylvania, indicating that its most critical aspects are a reasonable reflection of reality. Additional simulated effects resulting from varying mask usage, app adoption and initial number of infections match expected behavioral trends and further support confidence in the model.

Some initial conclusions can be drawn from experiments with SimAEN. Because of the widespread asymptomatic spread the most important aspect of any intervention relying on agent behavioral changes is adoption. Widespread masking is the most effective treatment because it reduces the transmission immediately, without the delay associated with AEN and MCT. These delays are a product of testing, but also the time that it takes for symptoms to develop in the generators. During this time there is ample opportunity for a large scale spreading event to occur.

MCT is particularly poor at reducing the number of cases. This is because of the low probability of an infectious agent identifying the agents they infected. Coupling this with the delays permitting transmission events means that the spread cannot be controlled through this method, provided that large scale spread is already occurring.

The best value to be achieved through AEN and MCT may come in the community surveillance that they permit instead of on their direct impact on the effective transmission rate. Public health could use this information to expand community testing efforts, direct messaging, and identify planned potential super-spreader events in these areas.

This study also indicates that AEN may have a difficult time finding a role in disease mitigation. When the spread rate is approximately at replacement ($R_{eff} \approx 1$), even a 75\% adoption rate has very minimal effect on reducing spread further. Even the small reduction it does afford comes at the cost of a substantial number of false quarantines. For high spread rates, AEN can reduce the spread rate at moderate adoption rates, but it involves the quarantining of large number of individuals.
\bibliography{references}{}
\bibliographystyle{plain}
\pagebreak
\section{Appendix: Parameter Default Values}
The table below presents default values for parameters of the SimAEN agent-based model. The values presented were derived from published literature, and from interviews with public health subject matter experts, when not available from literature.

%One of the keys to successful agent-based simulation is %identifying all of the influences and their appropriate values. %These parameters give each agent their unique characteristics %and define how the behave in each situation. Properly setting %these values for SimAEN required a combination of interviews %with subject matter experts and scouring of literature on the %topic. The tables below identify each of these parameters as %they occur by default.

\begin{table}[h]
\begin{tabular}{p{0.025\linewidth}p{0.025\linewidth}lp{0.15\linewidth}p{0.10\linewidth}}
\multicolumn{5}{l}{STARTING CONDITIONS} \\ \hline 
&\multicolumn{2}{p{0.65\linewidth}}{The number of cases at the start of the simulation} & 50000 & \cite{dong,cdc}\\ \hline 
\end{tabular}
\end{table}

\begin{table}[h]
\begin{tabular}{p{0.025\linewidth}p{0.025\linewidth}lp{0.15\linewidth}p{0.10\linewidth}}
\multicolumn{5}{l}{STOPPING CONDITIONS} \\ \hline 
&\multicolumn{2}{p{0.65\linewidth}}{Length of the simulation} & 30 days & \\
\rowcolor[HTML]{C0C0C0} &\multicolumn{2}{p{0.65\linewidth}}{Maximum number of current cases before program stops} & 1,500,000 & \\ \hline
\end{tabular}
\end{table}

\begin{table}[h]
\begin{tabular}{p{0.025\linewidth}p{0.025\linewidth}lp{0.15\linewidth}p{0.10\linewidth}}
\multicolumn{5}{l}{DISEASE PROPERTIES}\\ \hline 
&\multicolumn{2}{p{0.65\linewidth}}{The mean time between an individual being exposed and them becoming infectious} & 2 days & \cite{peirlinck,cdc2,johansson}\\ 
\rowcolor[HTML]{C0C0C0} &\multicolumn{2}{p{0.65\linewidth}}{The standard deviation of latent period} & 0.7 & \cite{peirlinck}\\ 
&\multicolumn{2}{p{0.65\linewidth}}{The mean time between an individual being exposed and them becoming symptomatic} & 6 days & \cite{peirlinck,cdc2,McAloon}\\ 
\rowcolor[HTML]{C0C0C0} &\multicolumn{2}{p{0.65\linewidth}}{The standard deviation of incubation period} & 2.3 days & \cite{peirlinck}\\
&\multicolumn{2}{p{0.65\linewidth}}{Infectious period} & 17 days & \cite{peirlinck}\\
\rowcolor[HTML]{C0C0C0} &\multicolumn{2}{p{0.65\linewidth}}{The likelihood an infected person will be asymptomatic} & 0.73 & \cite{poletti,cdc2}\\
&\multicolumn{2}{p{0.65\linewidth}}{The probability that a true contact event involving an infected person with no mask will result in infection...} &&\\
&&...if they are asymptomatic & 0.03&\\
&&...if they are presymptomatic & 0.03&\\
&&...if they are symptomatic & 0.07&\cite{laxminarayan}\\ \hline
\end{tabular}
\end{table}

\begin{table}[h]
\begin{tabular}{p{0.025\linewidth}p{0.025\linewidth}lp{0.15\linewidth}p{0.10\linewidth}}
\multicolumn{5}{l}{TESTING PARAMETERS} \\ \hline 
&\multicolumn{2}{p{0.65\linewidth}}{The probability that a person who has been called by public health will get tested on any given day} & 0.5 & \\ 
\rowcolor[HTML]{C0C0C0} &\multicolumn{2}{p{0.65\linewidth}}{The probability that a person who has no symptoms and has not been notified in any way will get a test} & 0.01&\\
&\multicolumn{2}{p{0.65\linewidth}}{The probability that a person who has received a notification through the app will get tested on any given day} & 0.5&\\
\rowcolor[HTML]{C0C0C0} &\multicolumn{2}{p{0.65\linewidth}}{The probability that a person who is symptomatic will get tested on any given day} & 0.5 & \cite{watson}\\
&\multicolumn{2}{p{0.65\linewidth}}{The mean and standard deviation of number of days that it takes for a test to get back (normal distribution)}&\\
&& Mean $\mu_{testing delay}$& 2 days & \cite{lazer}\\
&& Standard Deviation $\sigma_{testing delay}$& 1 day&\\
\rowcolor[HTML]{C0C0C0} &\multicolumn{2}{p{0.65\linewidth}}{Daily testing capacity} & $\infty$ &\\ \hline 
\end{tabular}
\end{table}

\begin{table}[h]
\begin{tabular}{p{0.025\linewidth}p{0.025\linewidth}lp{0.15\linewidth}p{0.10\linewidth}}
\multicolumn{5}{l}{PROBABILITY OF (+) TEST} \\ \hline
&\multicolumn{2}{p{0.65\linewidth}}{The probability that a person  will test positive given they are...}&&\\
&& ...exposed & 0.5&\\
&& ...presymptomatic & 0.75&\\
&& ...symptomatic & 0.9&\\
&& ...asymptomatic & 0.9&\\ \hline 
\end{tabular}
\end{table}

\begin{table}[h]
\begin{tabular}{p{0.025\linewidth}p{0.025\linewidth}lp{0.15\linewidth}p{0.10\linewidth}}
\multicolumn{5}{l}{APP PARAMETERS} \\ \hline 
&\multicolumn{2}{p{0.65\linewidth}}{The probability that a person is running the app} & 0.25, 0.5 &\\
\rowcolor[HTML]{C0C0C0} &\multicolumn{2}{p{0.65\linewidth}}{The probably that the phone of an infected person will exchange information with the phone of a close contact through Bluetooth detector settings (narrow, wide)} & 0.67, 0.86 &\\
&\multicolumn{2}{p{0.65\linewidth}}{The False Discovery Rate (FDR), used to create additional false positives picked up automatically by the system. At 0.5 the number of false discoveries will equal the number of true discoveries (narrow, wide)} & 0.23, 0.3 &\\ 
\rowcolor[HTML]{C0C0C0}&\multicolumn{2}{p{0.65\linewidth}}{The probability that a person who is running the app who gets a positive test will upload their key to public health} & 0.72 & \cite{wymant}\\
&\multicolumn{2}{p{0.65\linewidth}}{Is an individual required to be contacted by public health before uploading their key?} & No&\\ \hline 
\end{tabular}
\end{table}

\begin{table}[h]
\begin{tabular}{p{0.025\linewidth}p{0.025\linewidth}lp{0.15\linewidth}p{0.10\linewidth}}
\multicolumn{5}{l}{TRADITIONAL CONTACT TRACING (CT) PARAMETERS} \\ \hline 
&\multicolumn{2}{p{0.65\linewidth}}{The probability that a call from public health will reach a person identified through CT} & 0.5&\\
\rowcolor[HTML]{C0C0C0} &\multicolumn{2}{p{0.65\linewidth}}{The probability that a call from public health will reach a person expecting the call} & 0.75&\\
&\multicolumn{2}{p{0.65\linewidth}}{The probability that an exposed individual will be found using traditional CT} & 0.1&\\
\rowcolor[HTML]{C0C0C0} &\multicolumn{2}{p{0.65\linewidth}}{The maximum number of people an person can recall through traditional CT on a single phone call} & 10&\\
&\multicolumn{2}{p{0.65\linewidth}}{The number of contact tracers} & 500, 2,000 &\cite{website:pennsylvania, lewis}\\
\rowcolor[HTML]{C0C0C0} &\multicolumn{2}{p{0.65\linewidth}}{How long each contact tracer can spend on calling in a day} & 8 hours&\\
&\multicolumn{2}{p{0.65\linewidth}}{The number of times contact tracers will try to contact an individual before giving up} & 3&\\
\rowcolor[HTML]{C0C0C0} &\multicolumn{2}{p{0.65\linewidth}}{The length of time that a missed call takes} & 0.05 hours&\\
&\multicolumn{2}{p{0.65\linewidth}}{The length of time that a contact tracer takes to perform contact tracing on an index case} & 1 hour&\\
\rowcolor[HTML]{C0C0C0} &\multicolumn{2}{p{0.65\linewidth}}{The length of time that a public health call takes} & 0.1 hours&\\
&\multicolumn{2}{p{0.65\linewidth}}{The length of time it takes for a call to upload key} & 0.1 hours&\\ \hline 
\end{tabular}
\end{table}

\begin{table}[h]
\begin{tabular}{p{0.025\linewidth}p{0.025\linewidth}lp{0.15\linewidth}p{0.10\linewidth}}
\multicolumn{5}{l}{STARTING BEHAVIOR} \\ \hline 
&\multicolumn{2}{p{0.65\linewidth}}{The probability that a newly initialized individual will start in quarantine} & 0.0&\\ \hline 
\end{tabular}
\end{table}

\begin{table}[h]
\begin{tabular}{p{0.025\linewidth}p{0.025\linewidth}lp{0.15\linewidth}p{0.10\linewidth}}
\multicolumn{5}{l}{MASK PARAMETERS} \\ \hline 
&\multicolumn{2}{p{0.65\linewidth}}{The probability that a person will wear a mask before receiving EN, being called by contact tracer, developing symptoms or receiving a positive test} & 0.25, 0.5&\\
\rowcolor[HTML]{C0C0C0} &\multicolumn{2}{p{0.65\linewidth}}{The probability that a person will wear a mask while they are in the quarantine} & 0.9& \cite{Gurbaxani}\\
&\multicolumn{2}{p{0.65\linewidth}}{How much maskless transmission rate is proportionally reduced for each person wearing a mask (higher numbers mean less transmission risk)} & 0.65& \cite{ueki}\\ \hline 
\end{tabular}
\end{table}

\begin{table}[h]
\begin{tabular}{p{0.025\linewidth}p{0.025\linewidth}lp{0.15\linewidth}p{0.10\linewidth}}
\multicolumn{5}{l}{PERSONAL PARAMETERS} \\ \hline 
&\multicolumn{2}{p{0.65\linewidth}}{The probability that a person will call public health after a positive test} & 0.75&\\
\rowcolor[HTML]{C0C0C0} &\multicolumn{2}{p{0.65\linewidth}}{The probability that a person will call public health after receiving an EN notification} & 0.5&\\
&\multicolumn{2}{p{0.65\linewidth}}{Total number of people in a person's neighborhood (underlying Gaussian distribution for log-normal)}\\
&&Mean $\mu$& 2.5&\\
&&Standard deviation $\sigma$ & 1.1&\\
\rowcolor[HTML]{C0C0C0} &\multicolumn{2}{p{0.65\linewidth}}{The average number of contacts that an individual encounters each day if they take no precautions(underlying Gaussian distribution for log-normal)}&&\\
\rowcolor[HTML]{C0C0C0} &&Mean $\mu$& 2.9 & \cite{fu}\\
\rowcolor[HTML]{C0C0C0} &&Standard deviation $\sigma$& 1.0 & \cite{fu}\\
&\multicolumn{2}{p{0.65\linewidth}}{The average number of contacts that an individual encounters each day if they are in quarantine(underlying Gaussian distribution for log-normal)}&\\
&&Mean $\mu$ & 0.1&\\
&&Standard deviation $\sigma$ & 0.1&\\
\rowcolor[HTML]{C0C0C0} &\multicolumn{2}{p{0.65\linewidth}}{Probability of returning to starting behavior given negative test result and no symptoms} & 0.85&\\ \hline 
\end{tabular}
\end{table}

\begin{table}[h]
\begin{tabular}{p{0.025\linewidth}p{0.025\linewidth}lp{0.15\linewidth}p{0.10\linewidth}}
\multicolumn{5}{l}{PERSONAL BEHAVIOR} \\ \hline 
&\multicolumn{2}{p{0.65\linewidth}}{Probability of entering isolation given the person is symptomatic} & 0.9&\\
\rowcolor[HTML]{C0C0C0} &\multicolumn{2}{p{0.65\linewidth}}{Probabilities of entering isolation given the person receives a positive test} & 0.9&\\
&\multicolumn{2}{p{0.65\linewidth}}{Probability of entering quarantine given the person is successfully called by PH} & 0.75&\\
\rowcolor[HTML]{C0C0C0} &\multicolumn{2}{p{0.65\linewidth}}{Probability of entering quarantine given the person is notified by EN} & 0.5&\\ \hline 

\end{tabular}
\end{table}
\begin{table}[h]
\begin{tabular}{p{0.025\linewidth}p{0.025\linewidth}lp{0.15\linewidth}p{0.10\linewidth}}
\multicolumn{5}{l}{VACCINATION PARAMETERS} \\ \hline 
&\multicolumn{2}{p{0.65\linewidth}}{Probability that a person is vaccinated} & 0&\\
\rowcolor[HTML]{C0C0C0} &\multicolumn{2}{p{0.65\linewidth}}{Do vaccinated individuals spread disease asymptomatically?} & No&\\ \hline 
\end{tabular}
\end{table}

\end{document}